\documentclass[12pt]{article}
\usepackage{latexsym}
\usepackage{amsfonts}
\usepackage{amsmath, amsthm, amssymb}
\usepackage{graphicx}
\usepackage{fullpage}
\usepackage{setspace}
\usepackage{threeparttable}
\usepackage{longtable}
\usepackage{natbib}
\bibpunct{(}{)}{;}{a}{}{,}
\usepackage{dcolumn}
\newcolumntype{.}{D{.}{.}{-1}}
\newcolumntype{d}[1]{D{.}{.}{#1}}
\usepackage[top=1in, left=1in, right=1in,
bottom=1in]{geometry}
\usepackage{bm}
\usepackage[compact]{titlesec}
\usepackage{booktabs}
\usepackage{multirow}

\usepackage{color}
\usepackage{hyperref}

\usepackage{subfigure}

\usepackage{pgf,tikz}
\usetikzlibrary{positioning}
\usetikzlibrary{shapes}

\begin{document}
\pagestyle{plain}

\newcommand{\blind}{0}

\newcommand{\tit}{\bf Comparing Covariate Prioritization via Matching to Machine Learning Methods for Causal Inference using Five Empirical Applications}

\if0\blind

{\title{\tit\thanks{We thank Jake Bowers and Jas Sekhon for useful feedback on the analysis plan for this study.}}
\author{Luke Keele\thanks{Associate Professor, University of Pennsylvania, Email: luke.keele@gmail.com}
	\and Dylan S. Small\thanks{Professor, University of Pennsylvania, E-mail: dsmall@wharton.upenn.edu}
	}

\date{\today}

\maketitle
}\fi

\if1\blind
\title{\bf \tit}
\maketitle
\fi

\maketitle

\thispagestyle{empty}

\begin{abstract}
When investigators seek to estimate causal effects, they often assume that selection into treatment is based only on observed covariates. Under this identification strategy, analysts must adjust for observed confounders. While basic regression models have long been the dominant method of statistical adjustment, more robust methods based on matching or weighting have become more common. Of late, even more flexible methods based on machine learning methods have been developed for statistical adjustment. These machine learning methods are designed to be black box methods with little input from the researcher. Recent research used a data competition to evaluate various methods of statistical adjustment and found that black box methods out performed all other methods of statistical adjustment. Matching methods with covariate prioritization are designed for direct input from substantive investigators in direct contrast to black methods. In this article, we use a different research design to compare matching with covariate prioritization to black box methods. We use black box methods to replicate results from five studies where matching with covariate prioritization was used to customize the statistical adjustment in direct response to substantive expertise. We find little difference across the methods. We conclude with advice for investigators.

\end{abstract}

\clearpage

\doublespacing

\section{Introduction}

When attempting to use data for causal inference, randomized interventions are viewed as the ``gold standard'' due to the ability to remove measured and unmeasured confounding.  However, many interventions occur in settings where randomized experiments are difficult, impossible, or unethical. The primary alternative to an randomized trial is an observational study. \citet{Cochran:1965} defined an observational study as an empirical comparison of treated and control groups where the objective is to elucidate cause-and-effect relationships in contexts where it is not feasible to use controlled experimentation and subjects select their own treatment status. The primary weakness of observational studies is that when subjects select their own treatments, differences in outcomes may reflect initial differences in treated and control groups rather than treatment effects \citep{Cochran:1965,Rubin:1974}. Pretreatment differences amongst subjects come in two forms: those that have been accurately measured, which are overt biases, and those that are unmeasured, which are hidden biases. In an observational study, overt bias must be removed by investigators to render treated and control units comparable in terms of measured observed characteristics.

One approach for removing overt bias is to use statistical adjustment where investigators apply a method designed to removing overt bias to consistently estimate treatment effects. By far the most common method used for adjustment is some type of regression model \citep{Freedman:2005a}. However, regression models may suffer from bias due to functional form misspecification. To avoid this bias, a wide array of alternative forms of statistical adjustment have been developed. These more flexible methods include alternatives based on matching, weighting, or statistical learners (machine learning). Among these more flexible methods, different  philosophies have developed with respect to causal modeling.


Many of these methods--especially those based on matching--allow for some form of covariate prioritization so that subject matter expertise can be used to guide the removal of overt bias. In general, much of the recent matching literature has focused on adding functionality that allows more user input into the statistical adjustment process. See \citep{zubizarreta2012using} for the best example of this type of matching. Many methods based on machine learning take a very different approach. Here, the primary focus is on flexibly modeling the outcome. In general user input is considered unnecessary as an ensemble of statistical learners are applied. These differences in philosophy beg the question of whether investigators will be better off using subject matter expertise, or should they instead rely flexible data algorithms that need no input from experts?

Statistical simulation is the primary method used to compare and evaluate methods for statistical adjustment. Under this approach, data are generated where the truth is known and each method is applied repeatedly. The investigator can then measure comparative statistics on bias, coverage, or any other metric of interest. Comparisons based on simulation are widespread, since each method can be benchmarked against a known true quantity. Simulations, however, are ill-suited to understanding whether expertise should be used to guide methods of statistical adjustment.  While simulations can be designed to closely mimic real data structures, it is difficult to allow subject matter to guide a repeated simulation.

In this paper, we conduct a non-simulation based evaluation of relying on subject matter expertise using covariate prioritization compared to machine learning (ML) methods without user input. Our goal is to understand whether using subject matter expertise to guide statistical adjustment leads to fundamentally different results from ML methods that do not rely on subject matter expertise. While little current evidence exists on this point, one study, using simulation evidence, finds that ML methods are decisively better \citet{dorie2017automated}. In our research design, we identified five completed substantive applications where matching methods were customized to incorporate subject matter expertise. We then replicated the results of each study using machine learning methods that require no user input. We then compare the estimated treatment effects from these two different methods. We find that, in general, both methods tend to produce similar results. In only one instance do we find clear differences and that difference is easily explained as a result of using a particular covariate prioritization rather than matching per se. We conclude with a discussion of why both methods tend to agree in real data applications. Next, we review key assumptions.

\section{Review: Causal Identification via Conditional Ignorability}


Both customized forms of matching and ML adjustment methods are typically applied to observational data and invoke a set of causal assumptions that must hold for each method to produce consistent treatment effect estimates. Using the potential outcomes framework, we outline and review those assumptions. Let $Y_{i}$ denote the outcome and $Z_{i}$ denotes the treatment indicator such that we observe $(Y_{i}, Z_{i})$. For each individual $i$, let $Y_{i}(z)$ be the potential outcome given the treatment assignment value $z \in \{0,1\}$. Observed outcomes are related to potential outcomes in the following way: $Y_{i} = Y_{i} (Z_{i}) = Z_{i} Y_{i}(1)  + (1 - Z_{i}) Y_{i}(0)$. We denote observed covariates as $\mathbf{x}_{i}$, and $u_{i}$ is an unobserved covariate. Our notation implicitly assumes that there is no interference among units. This assumption is often referred to as one part of the stable unit treatment value assumption (SUTVA; \citealt{Rubin:1986}). We cannot estimate unit level causal effects, since we do not observe the potential outcomes. In general, investigators tend to focus on either the average treatment effect (ATE): 
\begin{equation}
ATE=\mathbb{E}\left[ Y_{i}(1) - Y_{i}(0) \right]
\end{equation}%
Alternatively, the average treatment effect may instead be defined for the subpopulation exposed to the treatment or the average treatment effect on the treated (ATT).
\begin{equation}
ATT=\mathbb{E}\left[ Y_{i}(1) - Y_{i}(0) \mid Z_i=1\right]
\end{equation}
Both quantities are forms of causal estimands, since both are based on a contrast of potential outcomes. For both estimands, investigators face an identification problem, since there are terms in the estimand that are unobservable. Even if we had samples of infinite size, one still could not estimate the average causal effect without observing both potential outcomes. As such, identification of these estimands requires additional assumptions in the form of an identification strategy. Researchers often use an identification strategy where treatment assignment is assumed to be strongly ignorable \citep{Rosenbaum:1983}. Under strong ignorability, the investigator assumes that treatment assignment is unconfounded
$$
\Pr(Z_{i} = 1 | \{ (Y_{i}(1), Y_{i}(0), \mathbf{x}_{i}, u_{i})\}) = \Pr(Z_{i} = 1| \{ \mathbf{x}_{i}\}),
$$
and probabilistic
$$
0 < \Pr(Z_{i} = 1|\{ (Y_{i}(1), Y_{i}(0), \mathbf{x}_{i}, u_{i})\}) < 1,
$$

Under this identification strategy, the analyst asserts that there is some set of covariates such that treatment assignment is as-if random conditional on these covariates. That is, treatment assignment is based on ``selection on observables'', where one assumes there are no unobservable differences between the treated and control groups \citep{Barnow:1980}. Critically, this assumption cannot be verified with observed data \citep{Manski:2007}, thus studies based on this assumption cannot ensure that some form of hidden bias may not be present. Researchers may check the plausibility of the selection on observables identification strategy by replicating estimates from randomize trials. The promise and the peril of using selection on observables is apparent in that it has been used to recover estimates from randomized trials in some cases \citep{Dehejia:1999,Sekhon:2005} but fails in other cases \citep{Arceneaux:2006,ramsahai2011extending}. However, under this set of assumptions, investigators must remove overt bias -- differences in covariates distributions across levels of $Z_{i}$.

Removing overt bias requires some type of statistical adjustment. Critically, even if selection on observables holds, treatment effect estimates can still be biased due to estimation error. We use the following equation from \citet{zhaocontest2018} which informally describes the sources of bias in any causal effect estimator:
\begin{equation*} \label{eq:decomposition}
\begin{split}
&~\text{Estimator} - \text{True causal effect} \\
=&~\underbrace{\text{Hidden bias}}_{\text{Due to design}}
~+~
\underbrace{\text{Misspecification bias}}_{\text{Due to modeling}}
~+~\underbrace{\text{Statistical
    noise}}_{\text{Due to finite sample}}.
\end{split}
\end{equation*}

Under selection on observables, hidden bias is assumed to be zero. Any remaining bias is due to either overt bias or modeling assumptions in the method of statistical adjustment. Traditionally, investigators have used regression models based on linear functional forms for statistical adjustment. If the relationship between confounders and the outcome or treatment status is nonlinear, bias may be present since the analysts relied on an overly restrictive method of statistical adjustment. Over the last twenty years, researchers have developed a variety of more flexible approaches including matching and propensity score weighting, which tend to have less misspecification bias. One critical question in the literature on causal inference focuses on the relative bias associated with different methods of statistical adjustment. Next, we review and contrast two forms of statistical adjustment. More specifically, we note a key philosophical distinction between matching methods and newer black box methods.

\section{Matching}

One method for removing overt bias is matching. Here, we provide a very brief review of the statistical literature on matching. We also focus on one recent strand of the matching literature, which has developed several techniques that allow for covariate prioritization in the matching process.

\subsection{Matching: A Review}

The concept of matching treated and control units with similar $\mathbf{x}_{i}$ distributions to remove overt bias has a long history in the statistics literature \citep{Rubin:1973b,Rubin:1973a} and is now well developed. Matching methods typically use optimization methods to form matched pairs with similar covariate distributions \citep{Rosenbaum:1989}. While pair matching is common, matched strata can take many forms depending on the study design \citep{Ming:2001,Ming:2000,Hansen:2006,Hansen:2004}. In general, matching first requires the calculation of a distance matrix that contains measures of covariate similarity between each treated unit and all controls. Propensity score distances and the Mahalanobis distance are frequently used to measure similarity between units. Matches are formed by using an algorithm to find treated to control assignments that minimize the total distances between two groups. See \citet{Sekhon:2013,Hansen:2004,Iacus:2011a,Rosenbaum:1989} for examples. Once matches are formed, investigators typically check how successful the matching process was by comparing treated and control covariate distributions. This diagnostics process is often referred to as balance checking. Matching is judged to be successful if overt bias is removed through the balancing of covariate distributions.

One important feature of matching is that the removal of overt bias happens without reference to outcomes. In fact, there are recommendations in the literature that all outcome data be withheld until matches are finalized \citep{Rubin:2008}. This occurs simply due to the mechanics of matching, but it is also consistent with a strand of the literature which emphasizes modeling the treatment assignment process rather than the outcome. One reason for this emphasis is that the withholding of outcome data until the matching is complete blinds the researcher to the treatment effect estimates when designing the study, avoiding bias from the researcher's preconceived opinions. However, one drawback of matching is that incorporating outcome information into the statistical adjustment process may improve the precision of the treatment effect estimate.

\subsection{Matching for Covariate Prioritization}

Next, we focus on the concept of covariate prioritization in matching.  Covariate prioritization occurs when the investigator decides to improve balance on a single covariate or set of covariates.  In general, prioritizing balance on one or more covariates comes at cost since that prioritization will tend to increase imbalances on other covariates. Prioritization comes at little cost, when the non-prioritized covariates are well balanced. As such, increasing imbalance on already well balanced covariates simply results in somewhat larger but still insignificant imbalances. Alternatively, the investigator may choose to ensure that one or a set of covariates are highly balanced, while other covariates remain imbalanced. Covariate prioritization is the primary means by which investigators can bring subject matter expertise to bear on statistical adjustment. That is, qualitative knowledge can guide the identification of which covariates should be prioritized. 

Covariate prioritization has long been possible in matching. The simplest form of covariate prioritization is exact matching. Here, treated and control units are matched exactly on a nominal covariate such as race. This serves as a form of covariate balance prioritization since all imbalance is removed for the covariate that is exactly matched. Caliper matching is another simple form of covariate balance prioritization when it is applied to specific covariates. Under a caliper match, two subjects are eligible to be matched if the covariate distance is less than a pre-specified tolerance \citep{Cochran:1973}. Some methods of covariate prioritization such as fine balance are computationally less costly variations on exact matching \citep{Rosenbaum:2007b,Yang:2012}.

However, more general forms of multivariate matching allow for more specific forms of prioritization. \citet{ramsahai2011extending} demonstrate how genetic matching can be used for highly flexible types of covariate prioritization. \citet{Pimentel:2015a} develops general method of covariate prioritization for standard optimal match methods based on network flows. Matching via integer programming allows for the most flexible form of covariate prioritization \citep{zubizarreta2012using}. This form of matching allows analysts to not only select covariates for prioritization, but also apply a wide variety of balance constraints such that investigators can very precisely select balance targets for covariates that have been prioritized.  While these methods are mostly used with matching, they can also be extended to methods that weight by the propensity score \citep{zubizarreta2015stable}.

\section{Black Box Methods}

The suite of statistical methods referred to as ``machine learning'' (ML) methods have typically been used for statistical prediction problems. However, recent work has adapted these methods to the estimation of treatment effects under strongly ignorability \citep{Mccaffrey:2004,Hill:2011a,Hill:2011,sinisi2007super,wager2017estimation}. As \citet{dorie2017automated} note, many of these ML methods are designed to be black-box approaches that require no input from the investigator.  As such, they stand in stark contrast to matching methods that are designed for the explicit incorporation of subject matter expertise via covariate prioritization. Next, we review two of these black box methods.

Bayesian additive regression trees (BART) is a nonparametric regression tree method for fitting arbitrary functions using the sum of the fit from many small regression trees. Like most ML methods, BART was originally designed for statistical prediction problems \citet{chipman2010bart}, but was one early example of a ML method being adapted for statistical adjustment under selection on observables \citep{Hill:2011a}. When BART is used for the estimation of treatment effects, it models the joint density of the treatment and covariates: $f(z,\mathbf{x}_{i})$.  This density is used to draw the posterior predictive distribution for $Y(1) = f(\mathbf{x}_{i},1)$ and $Y(0) = f(\mathbf{x}_{i},0)$. The empirical posterior distribution for the ATE or ATT is obtained by taking the differences between these quantities for each unit at each draw and averaging over the observations. In later work, \citet{Hill:2013} develops trimming rules for BART to allow it to better deal with a lack of overlap in the treated and control covariate distributions. \citet{Hill:2011} explicitly motivates BART as a black box method where user input is unnecessary and perhaps even undesirable. Moreover, BART focuses entirely on outcomes and does not model the treatment assignment mechanism. However, recent work has incorporated treatment assignment models into BART \citep{hahn2017bayesian}.

Another prominent ML method designed for treatment effect estimation under selection on observables is based on the concept of a Super Learner (SL) combined with Targeted Maximum Likelihood Estimation (TMLE)\citep{sinisi2007super,van2011targeted,gruber2015ensemble}. The SL approach is based on an ensemble algorithm. Here, the investigator can decide among a set of prediction methods--learners--that will then combined in the ensemble. The set of learners selected by the investigator are used to make out-of-sample predictions through cross-validation. The predictions from each learner are combined according to weights that minimize the squared-error loss from predictions to observations. These weights are then used to combine the fitted values from the learner when fit to the complete data. Then TMLE correction is applied to produce an estimate of the ATE or ATT. Under SL, the learners are used to model both the assignment mechanism as well as the outcome.

In sum, there are fairly clear philosophical differences in how best to implement statistical adjustment under selection on observables. One approach, that typically uses matching, envisions a role for subject matter expertise. Here, statistical adjustment is guided by qualitative input from substantive experts and expressed as prioritizing high levels of balance on a subset of covariates. Black box methods, in contrast, tend to have little role for such information and simply allow flexible fits to determine the final model. This contrast begs the question of which approach is optimal? Should investigators seek to incorporate substantive information into their adjustments or should they simply let black boxes do the work? Next, we focus on how we might answer this question.

\section{Research Design}

How might we judge which approach is best? Critically, there are significant challenges in even the design of a study that might compare the two approaches.
\citet{dorie2017automated} outline three specific challenges to designing studies focused on this question.  First, they note that the number of methods compared is usually limited. Second, the comparisons may not be calibrated to closely mimic features of real data applications. That is, simulated data generating processes may be overly simplified. Finally, many comparisons are subject to the file drawer effect, where comparisons are never published or made public when the results do not match expectations.

To overcome these limitations, \citet{dorie2017automated} designed a data analysis competition. In the competition, the organizers generated 77 simulated data sets that varied in terms of properties such a nonlinearity, ratio of treated to control observations, overlap, and other features. The data included 58 covariates not all of which were true confounders. Contestants could either submit treatment effect estimates and confidence intervals for 20 of the 77 data sets or submit an adjustment method that was applied to all 77 data sets to generate estimates and confidence intervals. Methods were judged on criteria such as bias, root mean-squared error and confidence interval coverage. Submissions were registered to avoid any results being suppressed. Contests were judged separately depending on whether they used just 20 data sets or all 77 data sets. In the competition, black box methods proved to be the best form of statistical adjustment. While the data analysis competition proved to be an innovative research design for comparing statistical adjustment strategies, some key weaknesses remained in terms of being able to compare black box methods to customized forms of matching. Primarily, even though the simulated data was based on a real data set, there was little role for substantive knowledge, since the data were presented to the competitors without identifying features. As such, there was little room for competitors to rely on substantive insights about features of the data.

We devised the following research design to directly compare statistical adjustments that rely on substantive expertise to black box methods. First, we identified a set of published or completed observational studies where covariate prioritization was applied and was guided by subject matter expertise.
We use the treatment effect estimates from these studies as benchmark estimates. Since these studies are completed, we did not alter the original statistical adjustments in any way. For each of the benchmark estimates from these studies, we performed a re-analysis using three different black box methods designed for the estimation of causal effects and saved the point estimates and 95\% confidence intervals. Below, we review how often the two approaches lead to substantively different conclusions. To avoid any appearance of bias in the selection of the studies, we published a pre-analysis plan which lists the five studies that were selected for replication using black box methods \citep{keelesmall2018}. The pre-analysis plan also included a summary of effect estimates for several of the studies based on the ATT when the ATT was not reported in the original study. In sum, for each of the five studies, we obtained the data. We then identified the exact set of covariates used in the published studies. Finally, we used the three ML methods to estimate the treatment effects from each study along with 95\% confidence intervals.

\subsection{Selected Box Methods and Comparison Criteria}

As outlined in \citet{keelesmall2018}, we selected three different ML methods to replicate the published studies. Specifically, we selected BART, SL methods, and generalized random forests (GRF) \citep{2016arXiv161001271A}.  All three are machine learning methods designed to estimate treatment effects with minimal input from the user. The primary weakness of our proposed research design is that we do not have a known true quantity to benchmark against. If every method produces a different result, we will have no way to know which, if any, of the methods is correct. However, our selection of ML methods was motivated by a strategy to help us identify why the treatment effect estimates may differ. We now outline this strategy. 

For each replication, there are four different possibilities in terms of the results. First, the point estimates and confidence intervals are not statistically distinguishable from each other. This result requires little explanation. Second, the point estimates are highly similar, but there are clear differences in the confidence interval lengths. This might occur since the ML methods use outcome information and matching does not. Third, we might find the point estimates have the same sign, but the confidence intervals do not overlap. Fourth, we might find that the point estimates have a different sign, and the confidence intervals do not overlap. Ideally, we would be able to trace exactly why the two final results happen. While formal characterization of why the methods produce different estimates is most likely not possible, we hope to provide insights into any differences via our choice of ML methods that we use for replicating the studies that used matching.

We selected the three ML methods to increase the possibility of identifying the source of any differences in the estimates. That is, each of the methods differ from each other in important respects. For example, while GRF is similar to BART in that it implements a flexible model for the outcome, GRF focuses on very local fits in the covariate space and BART fits a model that is global in the covariate space. BART only models the outcome, while the both the GRF and Superlearner approaches includes models for both the outcome and treatment assignment process. These differences may allow us to explain any differences we find across the replications. For example, if all the estimates based on blackbox methods differ from the matched estimate, this is probably due to the fact that the covariates selected for prioritization using subject matter expertise in the matches differ from the covariates given the most weight by the ML methods. We can explore this possibility by re-running the matches without covariate prioritization and observe whether the estimates move closer to those based on ML. However, if only one of the ML methods differ from the matched estimate, we will attribute the difference to the features of that ML method.

Next, we note a few specifics about the three black box methods we used. We used a Superlearner that was developed in \citet{kennedy2015semiparametric}. This implementation of SL has well defined asymptotic properties. We used an SL based on an ensemble of three learners: the generalized linear model, random forests, and the LASSO. BART, like matching, allows for trimming of treated units based on overlap in the covariate distributions. In our re-analyses using BART, we applied the trimming rule rules suggested in \citet{Hill:2013}. However, we found that estimates based on trimming did not differ from the estimates that did not trim the data. As such, we only report untrimmed estimates. For each of the ML methods, we did not attempt to perform any advanced tuning beyond the software defaults. Finally, since the primary causal estimand under matching is the ATT, we ensured that this was the target estimand for all three ML methods.

\section{Results}

Here, we present the results from all five replications. For each replication, we compare point estimates and 95\% confidence intervals. In one of the replications, estimation problems with one of the ML methods caused us to change the approach slightly. We detail those differences below.

\subsection{Study 1: Right Heart Catherization}

In the first application we selected for replication, the goal was to study the effect of Right Heart Catherisation (RHC) an invasive and controversial monitoring device that is widely used in the management of critically ill patients \citep{connors1996effectiveness}. \citet{ramsahai2011extending} used genetic matching methods to estimate the effect of RHC on mortality within 6 months, length of stay, and total medical costs. The investigators adjusted for the following set of covariates: sex, probability of 2--month survival estimated at baseline, coma score, an indicator for do not resuscitate status, the APACHE III acute physiology score, education, an index of daily activities 2 weeks prior to admission, Duke Activity Status Index, physiological measurements, ethnicity, income, insurance class, primary disease category, admission diagnosis, an indicator for cancer, PaO$_2$/FiO$_2$ ratio, creatinine, PaCO$_2$, albumin, number of comorbid illnesses, temperature, respiratory rate, heart rate, and white blood cell count. 

The investigators also sought to demonstrate how subject matter expertise can be incorporated into the statistical adjustment process. More specifically, they consulted a panel of clinical experts to identify a subset of covariates that should have higher priority in the matching process. Then they the prioritized balance on the covariates identified as most important by the panel of experts. The clinicians identified eight covariates that were of overriding importance, which included age, coma score, and the APACHE III acute physiology score. They altered the loss function in genetic matching \citep{Sekhon:2013} to give this set of covariates higher priority in the match. After matching, they reported average differences by treatment status on the three outcomes.

We replicated the results of this study using the three ML methods. Table~\ref{tab:rhc} contains the estimates from the original article as well as the estimates from the three ML methods. For the mortality outcome, the point estimates from matching and the three ML methods are all very similar. The confidence intervals based on the ML methods are narrower relative to matching, but differ little across the three methods. For the length of stay outcome, we find some minor differences. First, both BART and matching produce very similar results in terms of both the point estimate (1.25 vs. 1.4) and both confidence intervals include zero. The point estimates based on SL and GRF are larger and the confidence intervals no longer include zero as they are shorter. This pattern repeats itself with the cost outcome. BART and matching have similar point estimates, while the estimates based on SL and GRF are larger. The similarity between BART and matching is interesting given that matching focuses on the assignment mechanism, while BART models the outcome. In this setting, the confidence intervals based on matching are, however, notably longer than those based on the ML methods. 
 
\begin{table}[htbp]
\centering
\begin{threeparttable}
\caption{Outcome Analysis Comparison: Study 1 Right Heart Catherisation}
\label{tab:rhc}
\begin{tabular}{llccccc}
\toprule
 & & Mortality & Length of Stay & Total Costs \\
\midrule
\multirow{ 2}{1cm}{Match} & Point Estimate  & 0.046 & 1.25 & 9927  \\ 
 & 95\% Confidence Interval  & [ 0.00 , 0.09 ] & [ -1.16 , 3.66 ] & [ 3197 , 16,656 ] \\ 
\midrule
\multirow{ 2}{1cm}{BART} & Point Estimate  & 0.046 & 1.4 & 10496  \\ 
&95\% Confidence Interval  & [ 0.02 , 0.072 ] & [ -0.53 , 3.32 ] & [ 7495 , 13496 ] \\ 
\midrule
\multirow{ 2}{1cm}{SL} & Point Estimate  & 0.047 & 2.01 & 12985  \\ 
& 95\% Confidence Interval  & [ 0.025 , 0.068 ] & [ 0.6 , 3.41 ] & [ 10025 , 15944 ] \\ 
\midrule
\multirow{ 2}{1cm}{GRF} & Point Estimate  & 0.039 & 2.93 & 14986  \\ 
& 95\% Confidence Interval  & [ 0.014 , 0.064 ] & [ 1.42 , 4.44 ] & [ 11714 , 18259 ] \\ 
\bottomrule
\end{tabular}
\begin{tablenotes}[para]
{ \footnotesize Note: N = 5735 and 2184 were treated. Point estimates for mortality is difference in proportions. Length of Stay and Total Costs measured in days and dollars. BART: Bayesian Additive Regression Trees. SL: Superlearner. GRF: Generalized Regression Trees.} 
\end{tablenotes}
\end{threeparttable}
\end{table}

\subsection{Study 2: Minority Candidates and Co-Racial Turnout}

One area of study in political science focuses on identifying whether voter turnout is higher among minority populations when a co-racial candidates runs for office \citep{Barreto:2004,Brace:1995,Gay:2001,Griffin:2006,Tate:1991,Tate:2003,Voss:2001,Washington:2006,Whitby:2007,Fraga:2014,Fraga:2016,henderson2016cause,Keele:2011b}.\citet{Keele:2014d} study this question in the context of Louisiana mayoral elections from 1988 to 2011. In their study, voters are exposed to the treatment when at least one of the candidates on the ballot is African--American.  As such, the controls are elections with an all white slate of candidates. In the study, they matched municipalities with African--American mayoral candidates to municipalities without any African--American candidates for a given election. They matched cities and towns on population, the percentage of African--American residents, the percentage of residents with college degree, the percentage of residents with a high school degree, the percentage of residents unemployed, median income, the percentage of residents below the poverty line, an indicator for home rule municipal charter, and election year. 

The investigators used subject matter expertise to inform the statistical adjustments in a number of ways.  First, they identified the percentage of African--American residents in the municipality and year of the election as two key covariates. In the study, they set a balance constraint such that municipalities could differ by no more than a tenth of a percentage on percentage of African--American residents. Moreover, they enforced balance not only on central moment of this covariate but on higher moments as well. For the election year covariate, they used almost exact matching and allowed elections to differ by no more than two years. They performed three different matches: (1) for general elections, (2) for runoff elections, and (3) for a subset of runoff elections where it was thought that the threat of unobserved confounding was lessened. Given that the overlap between treated and control units was poor, the analysts applied optimal subset matching to find the largest set of observations that met the balance constraints. Optimal subsetting removed a significant number of treated observations. Note that because treated observations are trimmed, this implies that matching is no longer targeting the ATT. Instead, the match in this application is targeting a more local casual estimand. The outcome in the study was turnout among African--Americans in the municipality measured as a percentage.  

We replicated this study, using the three ML methods. To emulate the almost exact match on election year, we included year fixed effects in the ML specifications. Table~\ref{tab:bm} contains the treatment effect estimates for both the original study and those from the three black box methods. For the general elections, all three ML methods produce very similar results that are somewhat smaller than the estimate based on matching. Matching and SL produce confidence intervals that do not include zero, while BART and GRF have wider intervals that include zero. Interestingly, BART used all the original 1,006 data points, while the match is only based on 394 observations due to trimming. For the other two outcomes, there is little agreement across the four methods. In both cases, BART produces a result that largely agrees with matching. However, the estimate based on SL is negative, while the GRF estimates are close to zero. Unfortunately, all the confidence intervals are quite wide, making stronger conclusions difficult. Clearly small samples increase the difficulty of reliably producing estimates.

\begin{table}[htbp]
\centering
\begin{threeparttable}
\caption{Outcome Analysis Replication for Study 2: Minority Candidates and Co-Racial Turnout in Louisiana}
\label{tab:bm}
\begin{tabular}{llccc}
\toprule
& & General & Runoff & Runoff\\
& & Elections & Elections & Election Subset \\
\midrule
\multirow{ 2}{1cm}{Match} & Point Estimate & 3.41 & 5.11 & 2.97\\
& 95\% Confidence Interval &[0.72, 6.1] &[-3.5, 14] &[-11, 17]\\
& N  & 394 & 54 & 30\\
\midrule
\multirow{ 2}{1cm}{BART} & Point Estimate  & 2.69 & 4.59 & 5.35  \\ 
& 95\% Confidence Interval  & [ -0.42 , 5.8 ] & [ -0.5 , 9.69 ] & [ -0.51 , 11.2 ] \\ 
& N & 1006 & 187 & 96 \\ 
\midrule
\multirow{ 2}{1cm}{SL}  & Point Estimate  & 2.49 & -2.21 & -9.67  \\
& 95\% Confidence Interval  & [ 1.07 , 3.9 ] & [ -7.93 , 3.5 ] & [ -32.39 , 13.04 ] \\ 
& N & 1006 & 187 & 96 \\ 
\midrule
\multirow{ 2}{1cm}{GRF} & Point Estimate  & 2.27 & -0.23 & 0.76  \\
& 95\% Confidence Interval  & [ -2.07 , 6.61 ] & [ -6.77 , 6.32 ] & [ -6.53 , 8.06 ] \\ 
& N & 1006 & 187 & 96 \\ 
\bottomrule
\end{tabular}
\begin{tablenotes}[para]
{ \footnotesize Note: Point estimates are differences in turnout rates expressed as percentages. BART: Bayesian Additive Regression Trees. SL: Superlearner. GRF: Generalized Regression Trees.} 
\end{tablenotes}
\end{threeparttable}
\end{table}

\subsection{Study 3: Antibiotic Initiation in Critically Ill Children}

Study 3 is a study of the the effect of Procalcitonin (PCT)-guided antimicrobial stewardship protocols on antibiotic usage for patients admitted to a pediatric intensive care unit between 2011 and 2014 \citep{Ross:2017}.  In the study, infants placed on the PCT protocol were considered treated, infants that did not use the PCT protocol were controls. The investigators matched on age, race (African American yes/no), PRISM-III score, reason for PICU diagnosis, an indicator for chronic ventilator-dependent respiratory failure, an indicator for oncologic comorbidity, an indicator for new mechanical ventilation within the first hour of the PICU admission, source of PICU admission (7 categories), an indicator for surgery preceding PICU admission, and an indicator for trauma preceding PICU admission. 

The original study used two forms of covariate prioritization.  First, they included balance constraints on the higher moments for age and PRISM-III score.  This produced close matches along the entire distribution of these covariates instead of just for the mean. Second, patients were exactly matched on 26 diagnoses. The study focused on two outcomes: 1) a binary indicator for receiving less than 72 hours of therapy among patients for whom antibiotics were initiated and 2) a binary indicator for initiation of oral or parental antibiotic therapy in the 24 hours prior to or 48 hours after PICU admission. In the data, 505 infants were placed on the PCT protocol and 4,539 controls were available for matching.

Table~\ref{tab:pct} contains the original results along with the estimates from the three ML methods.  For the first outcome, all four methods produce very similar point estimates. Moreover, the confidence intervals are also quite similar. Here, the confidence intervals based on matching are not longer.  For the second outcome, matching, SL and GRF all return similar point estimates (0.20, 0.22, 0.22), with similar confidence intervals.  The length of the CI for matching is slightly longer (0.10 vs. 0.07). The estimate from BART is a slight outlier (0.29 vs 0.20), and the confidence interval only narrowly overlaps with the other methods. We suspect this discrepancy stems from the fact that BART places more emphasis on some global feature related to the outcome. 

\begin{table}[htbp]
\centering
\begin{threeparttable}
\caption{Outcome Analysis Replication for Study 3: Antibiotics in Critically Ill Children}
\label{tab:pct}
\begin{tabular}{llcc}
\toprule 
&  & Antibiotics $>$72 hours (0/1) & Antibiotic Initiation (0/1) \\
 \midrule
\multirow{ 2}{1cm}{Match} & Point Estimate & 0.112 & 0.202\\
& 95\% Confidence Interval &[0.049, 0.17] &[0.15, 0.25] \\
\midrule
\multirow{ 2}{1cm}{BART} & Point Estimate  & 0.09 & 0.29  \\ 
& 95\% Confidence Interval  & [0.04 , 0.15] & [0.23 , 0.34] \\
\midrule
\multirow{ 2}{1cm}{SL} & Point Estimate  & 0.11 & 0.22  \\ 
& 95\% Confidence Interval  & [ 0.06 , 0.15 ] & [ 0.184 , 0.248 ] \\ 
\midrule
\multirow{ 2}{1cm}{GRF} & Point Estimate  & 0.10 & 0.215  \\
& 95\% Confidence Interval  & [ 0.05 , 0.15 ] & [ 0.178 , 0.252 ] \\  
\bottomrule
\end{tabular}
\begin{tablenotes}[para]
{ \footnotesize Note: Point estimates are differences in proportions. N = 5,044 with 505 infants treated. BART: Bayesian Additive Regression Trees. SL: Superlearner. GRF: Generalized Regression Trees.} 
\end{tablenotes}
\end{threeparttable}
\end{table}

\subsection{Study 4: The 2010 Chilean Earthquake and Posttraumatic Stress}

The next replication uses the results from an investigation of the effect of the Chilean earthquake in 2010 on mental health \citep{zubizarreta2013designing}. The authors exploited the fact that a survey panel study that had been in the field before the earthquake, collected further data after the earthquake. In the study, Chilean residents that lived in areas directly affected by the earthquake were designated as treated units, and residents that lived far enough from the earthquake to be unaffected were designated controls. Residents that directly experienced the earthquake were paired to controls on 46 different covariates including age, gender, member of an indigenous ethnic group, household size, years of education, indicators for employment status, income, measures of housing status and measures of health status prior to the earthquake. The study design employed several form of covariate prioritization. The authors exactly matched on sex, whether a resident was a member of an indigenous ethnic group and age categories.  They also added fine balance constraints to scales of health and housing quality. Finally, balance constraints were applied to income to balance higher moments of the distribution. The study included a single outcome: the Davidson Trauma Scale (DTS). The DTS is a 17-item self-reported measure that assesses individuals for symptoms of post-traumatic stress disorder.

\begin{table}[htbp]
\centering
\begin{threeparttable}
\caption{Outcome Analysis Replication for Study 4: 2010 Chilean Earthquake and Posttraumatic Stress.}
\label{tab:equake}
\begin{tabular}{llcc}
\toprule 
& & Davidson Trauma Scale \\
\midrule
\multirow{ 2}{1cm}{Match} & Point Estimate & 18.4\\
& 95\% Confidence Interval &[17, 19]\\
\midrule
\multirow{ 2}{1cm}{BART} & Point Estimate  & 28.6  \\
&  95\% Confidence Interval  & [ 26.5 , 30.7 ] \\ 
\midrule
\multirow{ 2}{1cm}{SL} & Point Estimate  & 25.5  \\ 
&  95\% Confidence Interval  & [ 20.5 , 30.5 ] \\  
\midrule
\multirow{ 2}{1cm}{GRF} & Point Estimate  & 28.3  \\ 
& 95\% Confidence Interval  & [ 25.9 , 30.6 ] \\ 
\bottomrule
\end{tabular}
\begin{tablenotes}[para]
{ \footnotesize Note: Point estimates are differences in proportions. BART: Bayesian Additive Regression Trees. SL: Superlearner. GRF: Generalized Regression Trees. N = 10,057 with 2752 treated. Matched analysis based on sample size of 2520 matched pairs.} 
\end{tablenotes}
\end{threeparttable}
\end{table}

Table~\ref{tab:equake} contains the results from the matched analysis as well as the 3 ML methods. For the first time, we observe a clear difference between methods.  The matched estimate is 18.4 while the 3 ML methods have estimates that range from 25.5 to 28.6. and the confidence intervals do not overlap. Given the agreement between the ML methods, this would suggest that the black boxes placed greater weight on covariates that were not prioritized in the match. To explore this possibility, we ran a new match using the original data. For this match, we applied an optimal matching method with a robust version of the Mahalanobis distance and a propensity score caliper. For this match, we did not apply any specific balance constraints. The goal of this new match is to observe whether the balance constraints applied in the original match is the source of the discrepancy between the results. Table~\ref{tab:equake2} contains the outcome estimates from this new match. The treatment effect estimate from the new match is nearly identical to the estimates based on black box methods, and the confidence intervals overlap. Thus, if no covariate prioritization is employed, the estimates from matching agree with those based on ML methods  While we have traced the source of the discrepancy between the two methods, we do not know which is estimate is actually closest to the truth.

\begin{table}[htbp]
\centering
\caption{Outcome Estimates After Updated Match for Study 4: 2010 Chilean Earthquake and Posttraumatic Stress.}
\label{tab:equake2}
\begin{tabular}{lcc}
\toprule 
& Davidson Trauma Scale \\
\midrule
Point Estimate & 29.3\\
95\% Confidence Interval &[27, 32]\\
N  & 3842\\
\bottomrule
\end{tabular}
\end{table}


\subsection{Study 5: The Effectiveness of Emergency General Surgery in Elderly Patients}

We now turn to the final replication. In the original study, the authors sought to estimate the effectiveness of emergency general surgery (EGS) in adults over the age of 65 \citep{sharoky2017}. The investigators were also interested in whether the presence of a dementia diagnosis acted as an effect modifier for the effect of surgery. In the replication, we only focus on the main effect of surgery. 

To control for confounders, the authors matched on patient demographics, the number of comorbidities, an indicator for sepsis, an indicator for a pre-operative disability, a series of indicators for dementia type, indicators for 31 comorbidities, 9 indicators for broad medical condition types, 51 indicators for specific medical conditions, and two indicators for hospital admission source. The authors applied covariate prioritization by exactly matching on hospital, the 9 categories of surgical conditions, and the indicators for type of dementia.  They also applied near fine balance to the 51 more specific sub-condition indicators. In the study, they estimated treatment effects for the following set of outcomes: mortality (0/1), prolonged length of stay (0/1), discharge to a higher level of care (0/1), discharge to a hospice (0/1), and length of stay (in days). In our replication, we only focus on mortality, prolonged length of stay, and length of stay outcomes. The other two outcomes have patterns of missingness that cause additional complications. As in other replications, we include hospital fixed effects in the ML methods to mimic the exact matching on hospitals.

For this study, we were unable to produce BART estimates that were exactly analogous to the matched results. That is, we found that there were two aspects of the data that caused problems specifically for BART. The first problem we encountered is that we could not obtain BART estimates with the original sample size, which was 351,850 patients. BART required more memory (RAM) than was available on the secure computer where this data is housed.  Note this is the same computer used for the matching in the original paper and has 16 GB of RAM. As we noted above, dementia status was thought to be a possible effect modifier, and the investigators exactly matched on dementia status. To that end, we then applied BART just within the dementia subgroup, which had a much smaller sample size of 26,504. Again, BART could not allocate enough memory. We found that this was due to the large number of hospital fixed effects (more than 450).

To allow for a comparison across all the methods, we took the following approach. First, we re-ran the matching within the dementia subgroup and removed the hospital fixed effects from this match, which was done via exact matching on hospitals. This match has a smaller sample size and without the fixed effects should be feasible for estimation with BART. The outcome estimates based on the new match are nearly identical to the results from the original match. Table~\ref{tab:egs} contains the treatment effect estimates for the original match and for the smaller match. We then applied the 3 ML methods to the dementia subgroup and did not include hospital fixed effects. Table~\ref{tab:egs} contains the estimates from the 3 ML methods as well.

\begin{table}[htbp]
\centering
\begin{threeparttable}
\caption{Outcome Analysis Replication for Study 5: Emergency General Surgery in Elderly Patients}
\label{tab:egs}
\begin{tabular}{llccc}
\toprule
& & Prolonged LOS & Mortality & LOS\\
\midrule
\multirow{ 2}{1.1cm}{Original Match} & Point Estimate & 0.093 &-0.001 & 3.53\\
& 95\% Confidence Interval &[0.09, 0.095] &[-0.003, 0.0002] &[3.5, 3.6]]\\
\midrule
\multirow{ 2}{1.25 cm}{Modified Match} &  Point Estimate & 0.098 & -0.007 & 3.7\\
&  95\% Confidence Interval & [0.088, 0.11] & [-0.013, -0.0015] & [3.5, 3.9]\\
\midrule
\multirow{ 2}{1cm}{BART} &  Point Estimate  & 0.084 & -0.01 & 3.35  \\
&  95\% Confidence Interval  & [0.076 , 0.093] & [-0.014 , -0.005] & [3.24 , 3.46] \\
\midrule
\multirow{ 2}{1cm}{SL} &  Point Estimate  & 0.08 & -0.023 & 3.55  \\
&  95\% Confidence Interval  & [0.073 , 0.087] & [-0.031 , -0.016] & [3.37 , 3.73] \\ 
\midrule
\multirow{ 2}{1cm}{GRF} &  Point Estimate  & 0.082 & -0.018 & 3.64  \\
&  95\% Confidence Interval  & [0.075 , 0.09] & [-0.026 , -0.009] & [-0.19 , 0.16] \\  
\bottomrule
\end{tabular}
\begin{tablenotes}[para]
{ \footnotesize Note: Point estimates are differences in proportions for Prolonged LOS and Mortality. LOS is measured in days. Original N = 351850. Subset N = 26504 used in analysis.} 
\end{tablenotes}
\end{threeparttable}
\end{table}

In general, we find some minor differences between the ML treatment effect estimates and those based on matching. First, for the prolonged length of stay outcome, the ML estimates are all slightly lower than the estimates from matching. While the confidence intervals based on the Superlearner and matching do not quite overlap, the confidence interval from matching and the other two methods do overlap. In general, despite the slight differences, the methods provide highly similar results. For the mortality outcome, the estimates and confidence intervals based on BART and matching are very similar (-0.007 vs -0.01). The estimate based on the Superlearner, however, is larger and the confidence interval just narrowly fails to overlap with the confidence intervals for BART and matching.  While the GRF estimate is closer to the estimate from SL, the GRF confidence interval is wider such that it overlaps with all the other methods. Thus, we observe some differences in the estimates for the mortality outcome.  However, the differences are not large and do not vary by estimation method in an obvious way. Finally for the length of stay outcome, all the estimates are highly similar. We do not discern any clear pattern between methods and estimates. It is also worth noting that in this application use of outcome information under the ML methods typically provides no advantage in terms of narrower confidence intervals. For example, the length of the confidence interval for the prolonged length of stay outcome under matching is 0.022, while for BART the confidence interval length is 0.017.

\section{Discussion}

Here we outlined two different approaches to the analysis of observational data under a selection on observables identification strategy. Under one approach, black-box machine learning methods are applied with little guidance from the investigator.  Here, it is assumed that flexible fits are better than substantive knowledge.  Under the second approach, based on matching, substantive expertise can be incorporated in the statistical adjustment process 

In terms of general patterns, we note a few things. The most obvious pattern is that despite the underlying philosophical differences between these two approaches, they mostly produce nearly identical results.  In the one case where we found a clear difference, a more traditional match agreed with the black box estimates. Second, it is also the case that the black box methods at times produce shorter confidence intervals, but the differences are typically slight. Moreover, for larger sample sizes that advantage is very minimal.

At this point, we think it is worth pondering why there were so few differences in the results. Especially considering that black box methods performed so well in the contest described in \citet{dorie2017automated}.  For each study we analyzed, the hidden bias is assumed to be fixed--none of the methods we applied can reduce this bias. Given that there are so few differences in the estimates across methods, this suggests that the amount of misspecification bias is relatively small. Or at least, so small that matching is flexible enough to minimize any misspecification bias. This also suggests that the differences across methods in the contest is a result of a data generating process with large amounts of misspecification bias -- perhaps more misspecification bias than is present in most data sets.

Our replication process did reveal one clear computational advantage of matching. Under matching, once the match is completed relatively simply models can be fit for each outcome. For each of the black box methods, the entire method needs to be re-run for each outcome. For some of the larger data sets we worked with, there was a clear time-savings advantage to being able to run the match once and then quickly estimates treatment effects for each outcome. Also for the EGS replication, using the full data set, we were unable to produce BART estimates using a standard desktop computer, but were able to complete the matching 

Finally, we consider whether there are any larger lessons to be learned for the conduct of observational studies. We employed methods of statistical adjustment that appear to differ in fairly significant ways to a wide variety of data sets, and typically the treatment effect estimates were nearly identical. Thus one basic conclusion could be that so long as the method of adjustment is sufficiently flexible, analysts should employ whatever method they are most comfortable using. One might liken the situation to a setting where on average two medical procedures for treating a condition have similar average performance. A doctor might be best off using the procedure he or she is most comfortable with because the doctor will use it in a more skilled way.

Covariate prioritization methods and black box methods might be fruitfully used in conjunction.  As a first step, subject matter experts would be consulted to implement statistical adjustment with covariate prioritization applied.  In the second step, black box methods would be applied to the same data. If the results from the two methods agree, little else need be done. However, if the results differ, then investigators could explore what aspects of covariate prioritization produced different estimates. This would serve as useful way to explore whether conventional wisdom holds.

Finally, we think the dearth of differences serves to emphasize that in observational studies the form of statistical adjustment will rarely be what makes the evidence from a study compelling.  Instead aspects of study design are generally more important. For example, we would argue that the skillful use of negative controls or placebo tests \citep{lipsitch2010negative,Rosenbaum:2005} is far more likely to make result decisive than the use of matching or a black box method. \citet[ch. 5]{Rosenbaum:2010} offers one useful overview of devices that can make observational evidence more compelling. Alternatively, identification of a plausible natural experiment often provides another way to bolster the claims from observational studies \citep{Angrist:2010}.

\clearpage

\singlespacing
\bibliographystyle{/Users/ljk20/Dropbox/texmf/bibtex/bst/asa}
\bibliography{/Users/ljk20/Dropbox/texmf/bibtex/bib/keele_revised2}

\end{document}